\documentclass[aps,amsfonts,pre,twocolumn,showpacs,multicol]{revtex4-1}
\usepackage{epsfig,amsmath,amssymb,bm,epsf,graphics,psfrag,verbatim,subfigure,bm,hyperref}

\def\st{\epsilon} % The strain tensor
\def\numconst{N_c} % The number of bonds/constraints/springs in the system
\def\numzero{N_0} % the number of zero modes
\def\numss{N_{ss}} % the number of self stresses
\def\numdof{N_{\textrm{d.o.f.}}} % The number of degrees of freedom
\def\tstb{\tilde{\boldsymbol{\epsilon}}}
\def\tst{\tilde{\epsilon}}
\def\D{\det{\tstb}}

\def\TTMM{transformable topological mechanical metamaterials}
\def\TTMMab{TTMM}

\def\fr{\phi}

\begin{document}
\title{Transformable topological mechanical metamaterials}

\author{D. Zeb Rocklin}
\affiliation{Department of Physics, University of Michigan, Ann Arbor MI 48109-1040}
\author{Shangnan Zhou} 
\affiliation{Department of Physics, University of Michigan, Ann Arbor MI 48109-1040}
\author{Kai Sun}
\affiliation{Department of Physics, University of Michigan, Ann Arbor MI 48109-1040}
\author{Xiaoming Mao}
\affiliation{Department of Physics, University of Michigan, Ann Arbor MI 48109-1040}
\date{\today}

\date{\today}

\begin{abstract}
Mechanical metamaterials are engineered materials that gain their remarkable mechanical properties, such as negative Poisson's ratios, negative compressibility, phononic bandgaps, and topological phonon modes, from their structure rather than composition.  
Here we propose a new design principle, based on a uniform soft  deformation of the whole structure, to allow metamaterials to be immediately and reversibly transformed between states with contrasting mechanical and acoustic properties.  These properties are protected by the topological structure of the phonon band of the whole structure and are thus highly robust against disorder and noise.  We discuss the general classification of all structures that exhibit such soft deformations, and provide specific examples to demonstrate how to utilize soft deformations to transform a system between different regimes  such that remarkable changes in their properties, including edge stiffness and speed of sound, can be achieved.

\end{abstract}
\maketitle

If a material has the ability of tuning its mechanical properties, such as stiffness, in real time, there will be broad potential applications.  
%The ability to tune mechanical properties, such as stiffness, of a material has broad potential applications.   
%In traditional materials, mechanical properties, such as stiffness, can hardly be tuned once the material is made.  
%``Smart'' materials whose properties can be easily altered in a controlled way in response to environmental cues have broad technological applications~\cite{Smith2005,Florijn2014,Silverberg2014,Wang2014,Grima2013,Greaves2011,Lakes2008,Nicolaou2012,Kushwaha1993}. 
For example, %in the exploration of space, 
we can imagine a reusable launch system for space exploration made of such a material, 
where the space vehicle is rigid during takeoff and in orbit, but during landing the rigid surface of the vehicle transforms 
into a soft cushion layer to absorb the impact.
However, this is highly challenging, because stiffness is an intrinsic property.  Traditional techniques to change stiffness of a material are either irreversible, e.g., photo-polymerization that dentists use to rigidify dental fillings, or involve significant stress in the material, e.g., tightening a guitar string.  It is not until recently there have been proposals of mechanical metamaterials with tunability~\cite{Kornbluh2004,Florijn2014,Silverberg2014,Wang2014,Grima2013,Greaves2011,Lakes2008,Nicolaou2012,Eidini2015}.

\begin{center}
\begin{figure*}[t]
\includegraphics[width=.9\textwidth]{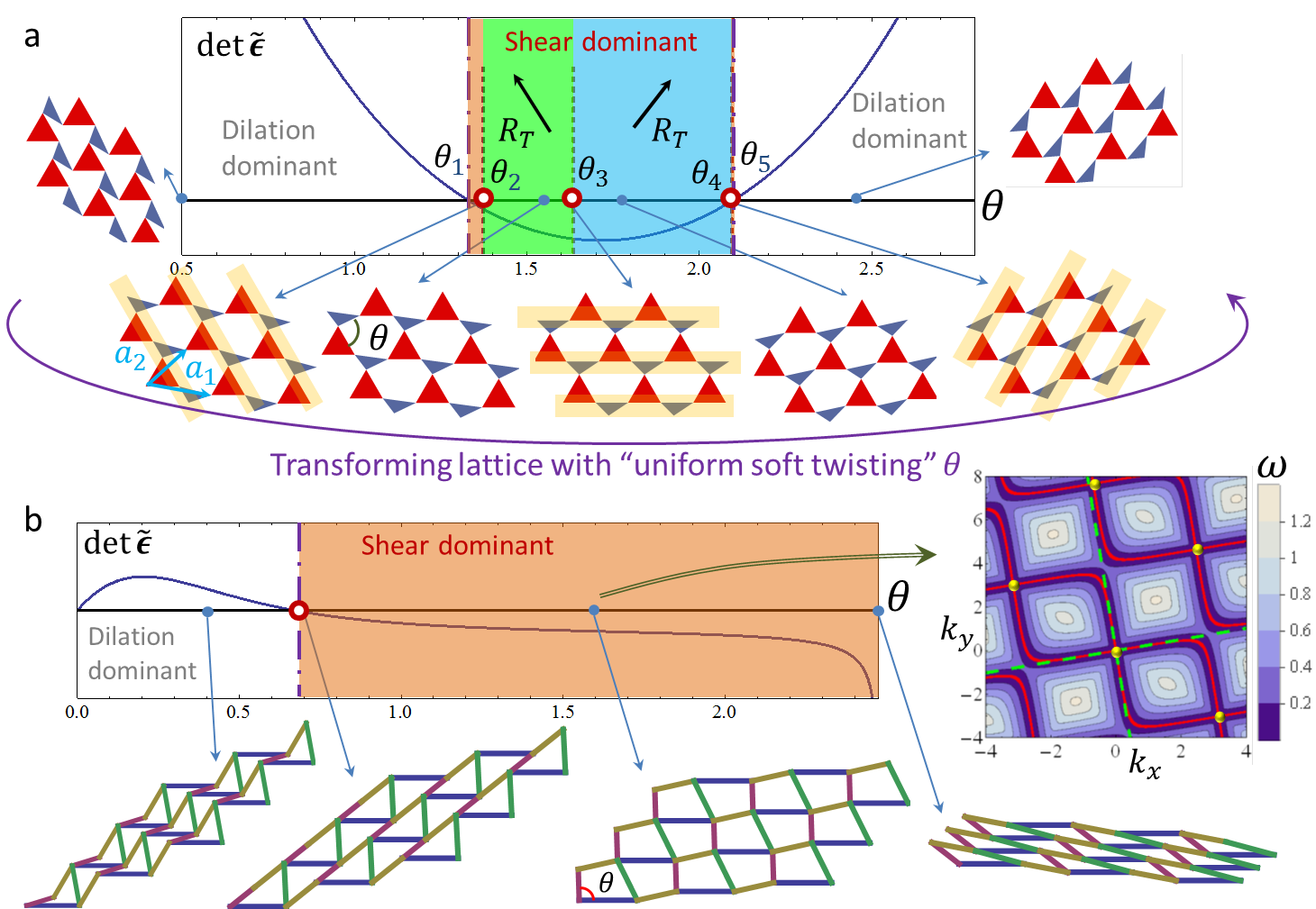}
\caption{
(a) Uniform soft twisting of a deformed kagome lattice.  Two types of triangles (red and blue) are connected by free hinges at their corners, forming a deformed kagome lattice with primitive vectors $\mathbf{a_1}, \mathbf{a_2}$.  The angle $\theta$ between the triangles defines the twisting coordinate.  The blue curve shows $\D$ [defined in Eq.~\eqref{EQ:epsilon}] as a function of $\theta$.  
The 3 white dots on the $\theta$ axis represent 3 critical angles ($\theta_2,\theta_3,\theta_4$) where sides of the triangles form straight lines (yellow stripes on the lattices) and topological polarization $\mathbf{R}_T$ (shown as black arrows above the axes) changes.  (b) Uniform soft twisting of a deformed square lattice constructed of 4 struts of different lengths (4 different colors), with each primitive unit cell contains 2 hinges.  The spatially varying floppy modes in the deformed square lattice when $\D<0$ are bulk instead of floppy edge phonons.  An example of the phonon frequency contour plot (as a function of momenta $k_x, k_y$) is shown on the right, where the zero frequency phonon modes are shown in red, the two green dashed lines show the two zero speed of sound directions $(1, \lambda_{\pm})$ given by Eq.~\eqref{EQ:lambda}, and the yellow dots show reciprocal lattice sites.
}
\label{fig:twisting}
\end{figure*}
\end{center}

In this Article, we propose a new design principle for smart mechanical metamaterials, which we name ``\TTMM'' (\TTMMab), whose mechanical and acoustic properties can be easily tuned by orders of magnitudes
without the need to disassemble/reassemble the system.  Our design utilizes soft deformations, also known as {\it mechanisms} or {\it floppy modes}, 
which change configurations of the material with little %or no 
energy cost.  Structures with floppy modes are ubiquitous in natural and 
engineered systems, e.g., a synovial joint in human body or a door hinge. Our design, as shown in Fig.~\ref{fig:twisting}, involves periodic 
structures consisting of rigid building blocks (polygons or struts) connected by flexible hinges, which can be created using existing technologies, like 3D printing~\cite{Paulose2015a} or self-assembly~\cite{Pelesko2007,Mao2013,Mao2013a,Rocklin2014}. 
These structures can exhibit soft deformations involving changing angles between building blocks at hinges without deforming any building blocks.  
These soft deformations are either uniform, i.e., all repeating units twist in the same way (soft deformations of this type has been called ``Guest modes''~\cite{Guest2003,Lubensky2015}), or spatially varying, i.e., blocks at different locations show different twistings. 
As shown in the 
\href{http://www-personal.umich.edu/~maox/research/TTMM/TTMM.html}{Video in the Supplementary Information (SI)} ~\cite{TTMMVideo}
%video in the Supplementary Information (SI), 
, the uniform soft deformations, which we name ``uniform soft twistings'', can be easily manipulated by a simple 
expansion of the lattice, and they serve as tuning knobs that control the mechanical properties of the system, transforming the edge of the system from soft to rigid. 

Our design principles for the \TTMMab\ are based on the following findings. First, utilizing elastic theory, we prove that if a two-dimensional (2D) structure exhibits one uniform soft twisting, a series of spatially varying floppy modes must also exist. 
A proof of this using the linearized strain tensor has been given in Ref.~\cite{Lubensky2015}, and here we show that the same general conclusion holds for fully nonlinear strain.  
Then, using this theorem, we find that all 2D structures with uniform soft twistings can be classified into two categories: dilation dominant and shear dominant. Systems in these two regimes show very different elastic properties.  In the dilation dominant 
regime, the bulk of the material is a rigid 2D solid, but all edges are soft due to floppy modes that reside on edges~\cite{Sun2012}. In the shear dominant regime, however, 
floppy modes arise in the bulk, while the properties of the edges can differ sharply. Depending on the architecture, some of the edges may become rigid, 
while others remain soft. Finally, we further
show that these two different regimes can be realized in the \emph{same} mechanical structure: the soft
twisting can reversibly shift it from one regime to the other and hence
alter various properties, including edge stiffness and sound speed by
orders of magnitude.
%%%%%%%%%%%%%%%%%%%%%%%%%%%%%%%%%%%%%

In addition to this controllability, mechanical properties of these systems show extraordinary robustness.  
%In addition to controllable mechanical properties, these structures show extraordinary robustness.  
For example, as mentioned above, a system in the dilation dominant regime displays rigid bulk and soft surface. 
Although similar mechanical properties can be achieved using composite materials, e.g., by covering a stiff solid (e.g., a metal) with a soft cushion layer (e.g., rubber), there is one key difference between the two: robustness.  
The composite structure lacks robustness, i.e., if the outer cushion layer peels off, the protection will be gone. However, the soft surface layer always exists in the smart mechanical metamaterial we design.  
Because the whole structure is built from the same building blocks, after the outer layers peels off, the newly exposed surfaces will \emph{become soft}.  
Such robustness is of critical importance for devices working under severe conditions, whose protecting layer may wear out due to extreme temperature or friction, etc.

The origin of this extraordinary robustness lies in \emph{topological protection}.  As we will explain below, in addition to control the elastic properties,
in certain systems, uniform soft twistings can also trigger topological transitions,  where phonon modes (sound waves) change their topological structure.
%in the two regimes exhibit different topology. 
%Such topological phonon structures Topological structure for mechanical systems was first discovered in Ref.~\cite{Kane2014}, and is analogous to electronic topological states of matter, 
As discovered recently, mechanical systems may exhibit different topologies~\cite{Prodan2009,Kane2014,Chen2014,Vitelli2014,Paulose2015,Chen2015,Paulose2015a,Xiao2015,Xiao2015a,Po2014,Yang2015,
Nash2015,Wang2015,Wang2015a,Susstrunk2015,Kariyado2015,Peano2015,Mousavi2015,Khanikaev2015}, in strong analogy to topological states
in electronic systems, e.g., topological insulators \cite{Hasan2010,Moore2010,Qi2011}. 
One key feature of topological states is the {\it bulk-edge correspondence}, in which edge properties of a topological system are 
dictated by the bulk of the system, and are independent of surface conditions or microscopic details.
This bulk-edge correspondence offers strong protections to the edge properties in these systems  against noise and local perturbations, 
e.g., peeling off surface layers or inserting impurities. The only possible way to modify the edge properties 
%%%%%%%%%%%%%%%
is by changing the topological structure of the whole system, which requires
changing globally the entire structure, e.g., a uniform soft twisting discussed above.
%This is the root of the robustness in these systems and the control of their elastic properties.
This is the root of the robustness in these systems and why we can control their elastic properties through soft twisting.
%To highlight the role of topology, which works behind the scene for our designs, we name our proposed mechanical metamaterial ``\TTMM'' (\TTMMab).

\noindent\textbf{Elastic theory and general classification}\\
We start the analysis by considering an arbitrary 2D elastic system that exhibits (at least) one uniform soft twisting, 
while how to design such a structure will be discussed later.
Generally,  deformations of an elastic medium can be described using the left Cauchy-Green strain tensor.
Here, we use $\tstb$ to denote the strain tensor for the uniform soft twisting 
\begin{align}\label{EQ:epsilon}
	\tstb  =\begin{pmatrix}
	\tst_{xx}& \tst_{xy} \\
	\tst_{xy} &  \tst_{yy}
	\end{pmatrix},
\end{align}
which is independent of position.  
%As shown in the SI, the fact that this deformation cost zero elastic energy immediately implies that there must exist two families of spatially varying floppy modes
As proved in the SI, utilizing the fact that any elastic deformation in flat space must have zero curvature, 
the existence of the zero energy uniform deformation $\tstb$ leads to 
%Because $\tstb$ costs no elastic energy, there must exist 
two families of spatially varying floppy modes
described by strain tensors
\begin{align}\label{EQ:fm}
	&\boldsymbol{\st}_{+}(\mathbf{r})=\tstb \; f_{+}(x+\lambda_{+} y), \nonumber\\
	&\boldsymbol{\st}_{-}(\mathbf{r})=\tstb \; f_{-}(x+\lambda_{-} y)
\end{align}
where $\mathbf{r}=(x,y)$ is the coordinate, $f_{\pm}(w)$ are two arbitrary scalar functions and $\lambda_{\pm}$ are two constants determined by $\tstb$ 
\begin{align}\label{EQ:lambda}
	\lambda_{\pm}=(\tst_{xy}\pm\sqrt{-\D})/\tst_{xx},
\end{align}
where $\D=\tst_{xx}\tst_{yy}-(\tst_{xy})^2$ is 
the determinant of $\tstb$.

As shown in the SI,  the elastic energy of these floppy modes ($E$) vanishes at the leading order, i.e., $E\sim O(\epsilon^3)$, 
which is much lower than that of a typical elastic deformation with $E\sim O(\epsilon^2)$.
This is why they are dubbed as floppy modes or soft modes. 
%As will be discussed below, for a certain family of systems these floppy modes cost exactly zero elastic energy due to the protection of a counting rule.

These floppy modes lead to the existence of sound waves with zero velocity, i.e., soft phonon modes.
It is known that the speed of sound is proportional to the square root of the
corresponding elastic constant. Here, because our floppy modes have vanishing elastic energy to $O(\epsilon^2)$, 
their corresponding elastic constants and sound velocities vanish. 
Below, we show that these soft phonon modes can be either bulk or edge phonon modes.
%, and their frequency can be either exactly zero or proportional to $k^2$ ($k$ is the wave number).
\begin{center}
\begin{figure*}[t]
\includegraphics[width=.8\textwidth]{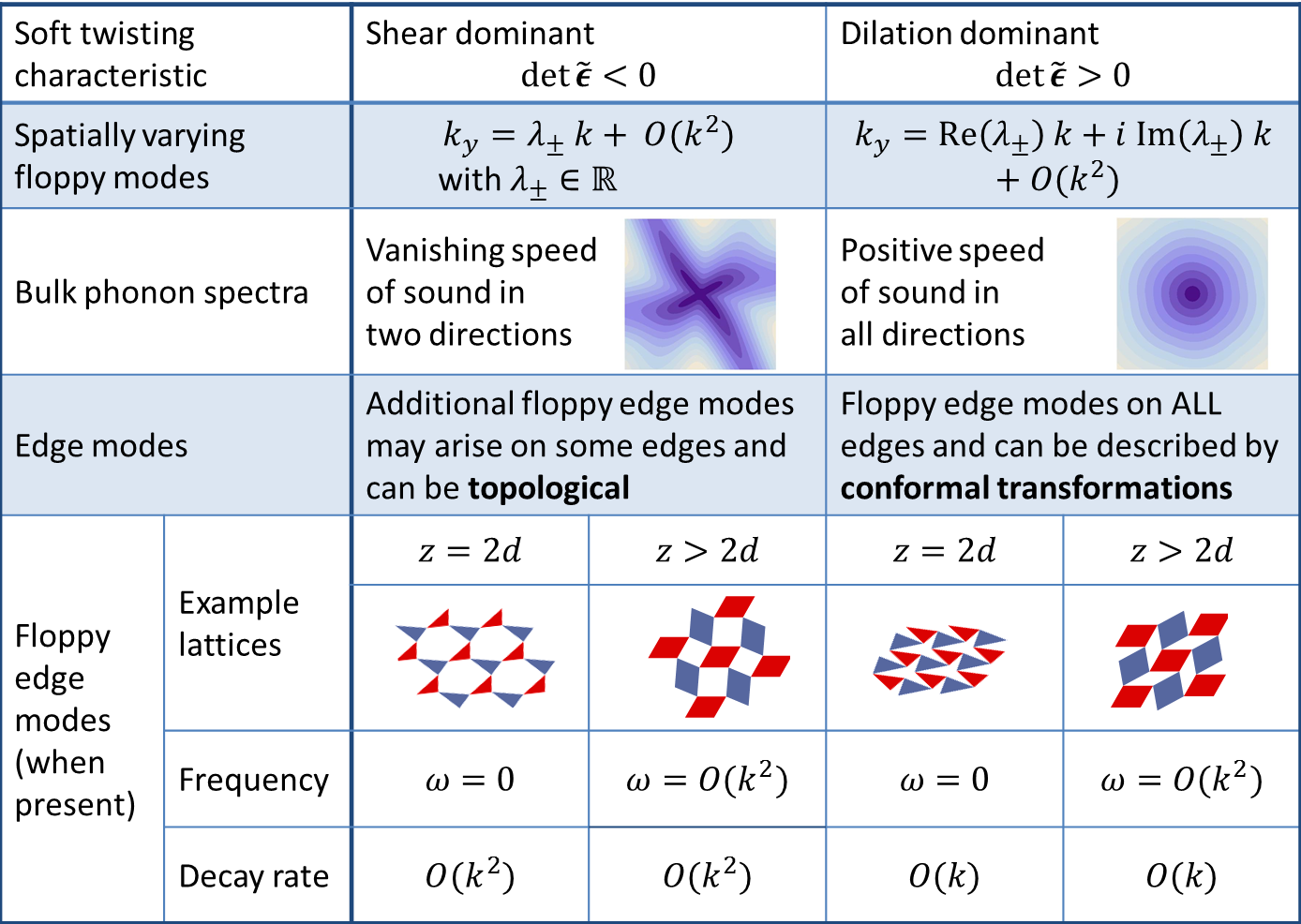}
\caption{
General classification of lattices with uniform soft twistings.  The spatially varying floppy modes are expressed in terms of the wave number in the $y$ direction when a plane wave of wave number $k$ propagates in the $x$ direction [see discussions after Eq.~\eqref{eq:surface_mode_general}].  The bulk phonon spectra show example phonon frequency contour plots as a function of $k_x, k_y$.  The example lattices are shown as rigid polygons (triangles or parallelograms) connected by free hinges at their corners~\cite{Grima2005}, and they can be directly mapped into strut-hinge frames by replacing the triangles by 3 connected struts on their edges and the parallelograms by 5 connected struts with 4 on edges and 1 on the diagonal to make it rigid.  Thus the structure consists of triangles (deformed kagome lattices as defined the text) have $\langle z \rangle=4=2d$ and the structure consists of parallelograms (deformed checkerboard lattice) have $\langle z \rangle=5>2d$.
}
\label{fig:table}
\end{figure*}
\end{center}

The characteristics of these floppy modes are dictated by the sign of $\D$, which distinguishes two different regimes:
the dilation dominate regime, $\D>0$ and the shear dominate regime, $\D<0$. 
In general, the uniform soft twisting may contain both dilation ($\tilde{\epsilon}_{xx}$ and $\tilde{\epsilon}_{yy}$) 
and shear deformation ($\tilde{\epsilon}_{xy}$). If the uniform soft twisting is dominated by dilation (shear), 
we have $\tilde{\epsilon}_{xx}\tilde{\epsilon}_{yy}>\tilde{\epsilon}_{xy}^2$ 
($\tilde{\epsilon}_{xx}\tilde{\epsilon}_{yy}<\tilde{\epsilon}_{xy}^2$), which gives a positive (negative) $\D$.
It is worthwhile to emphasize here that $\D$ measures the intrinsic property of the uniform soft twisting and it is independent of 
the choice of coordinates.  In addition, structures in the dilation dominant regime are necessarily auxetic~\cite{Greaves2011} because they have $\tilde{\epsilon}_{xx}\tilde{\epsilon}_{yy}>0$ which gives a negative Poisson's ratio.

In the dilation dominant regime ($\D>0$), the floppy modes are edge modes confined to %localized on
all edges of the system. 
This conclusion is transparent after we decompose the two arbitrary functions $f_{\pm}$ into Fourier 
series $f_{\pm}(w)=\sum_k \phi_{\pm}(k) e^{i k w}$ so that the functions in Eq.~\eqref{EQ:fm} turn into
\begin{align}\label{eq:surface_mode_general}
	f_{\pm}(x+\lambda_{\pm} y)=\sum_{k} \phi_{\pm}(k) e^{i k x + i \lambda_{\pm} k y}.
\end{align}
For any real number $k$, along the $x$ direction, the exponential factor $e^{i k x}$ describes a plane wave with wave number $k_x=k$. 
However, along $y$, because $\lambda_{\pm}$ is complex for $\D>0$, its imaginary part, $ \textrm{Im}\lambda_{\pm}$, yields 
a factor $e^{-\kappa y}$ with $\kappa=k \,\textrm{Im}\lambda_{\pm}$, 
so that the amplitude of this deformation decays exponentially along the $y$ axis. If the system has an open edge parallel to the $x$-axis,
this is a plane wave along the edge whose amplitude decays exponentially from the edge into the bulk of the system, i.e., an edge mode with zero sound velocity.
The decay rate for this edge mode is proportional to the wavevector, $\kappa \propto k$.
Because the $x$-direction here is chosen arbitrarily, the same conclusion applies to arbitrary edge directions and thus 
floppy modes arise on all edges. Because the elastic theory shows no bulk floppy modes, the bulk is in general rigid and has no floppy mode except the uniform soft twisting, which we assumed from the beginning.
One special case in the dilation dominant regime, the twisted kagome lattice, was discussed in Ref.~\cite{Sun2012}, where the uniform soft twisting is a pure dilation 
$\tilde{\epsilon}_{xy}=0$ and $\tilde{\epsilon}_{xx}=\tilde{\epsilon}_{yy}$. 
For that special case, the system has an emergent conformal symmetry and the floppy edge modes
are conformal deformations. As we prove here, the same qualitative properties shall always arise 
as long as  $\D>0$.

For the shear dominant regime $\D<0$, the floppy modes are bulk plane waves along two special directions. 
This can be seen directly from Eq.~\eqref{eq:surface_mode_general}. With negative $\D$, 
$\lambda_{+}$ and $\lambda_{-}$ are both real, 
and thus $f_{+}$ and $f_{-}$ both describe bulk plane waves along the two directions of $k_y=\lambda_+ k_x$ and 
$k_y=\lambda_- k_x$.
For bulk sound waves along these two special directions, the sound velocity vanishes, which is the key signature of the shear dominant regime.
On the edge of the system, our general elastic theory neither requires nor prevents the existence of floppy edge modes, 
implying that the fate of the edge is not universal and relies on the architecture of the lattice. 
Generally in a solid, surface or edge sound waves, known as Rayleigh waves, could arise and the 
frequencies of these Rayleigh waves are lower than those of waves in the bulk (surface waves can also have frequencies located 
in a phonon band gap, but because we only focus on low-frequency phonon modes, 
this case will not be considered here)~\cite{Landau1986}.
For our structures with uniform twistings, similar Rayleigh waves may arise for certain edges.
Because their frequencies  are lower than the bulk ones, including the floppy bulk plane waves with zero sound velocity, 
these surface waves shall also be soft and have zero sound velocity. 
At long wave lengths (small $k$), these floppy edge modes have decay rate $\kappa\sim k^2$ and penetrate much deeper into the bulk, 
in comparison to the floppy edge modes in the dilation dominant regime discussed above, which has $\kappa\sim k$. 

Finally, it is worthwhile to point out that the above general discussions are all based on the existence of a soft uniform deformation $\tst$, without assuming any microscopic structures.  Knowledge of the microscopic structures provides more information on what form these floppy modes take.  
In particular, in periodic structures built from struts and flexible hinges, as we discuss later, if the structure satisfies the Maxwell lattice condition $\langle z \rangle =2d$ (where $\langle z \rangle$ is the mean number of struts connecting to one hinge and $d$ is the spatial dimension), the aforementioned floppy modes may become of exactly zero energy.  In contrast, this is not guaranteed in more connected structures with $\langle z \rangle >2d$.  We summarize this classification in the table in Fig.~\ref{fig:table}.

\begin{center}
\begin{figure*}[t]
\includegraphics[width=.9\textwidth]{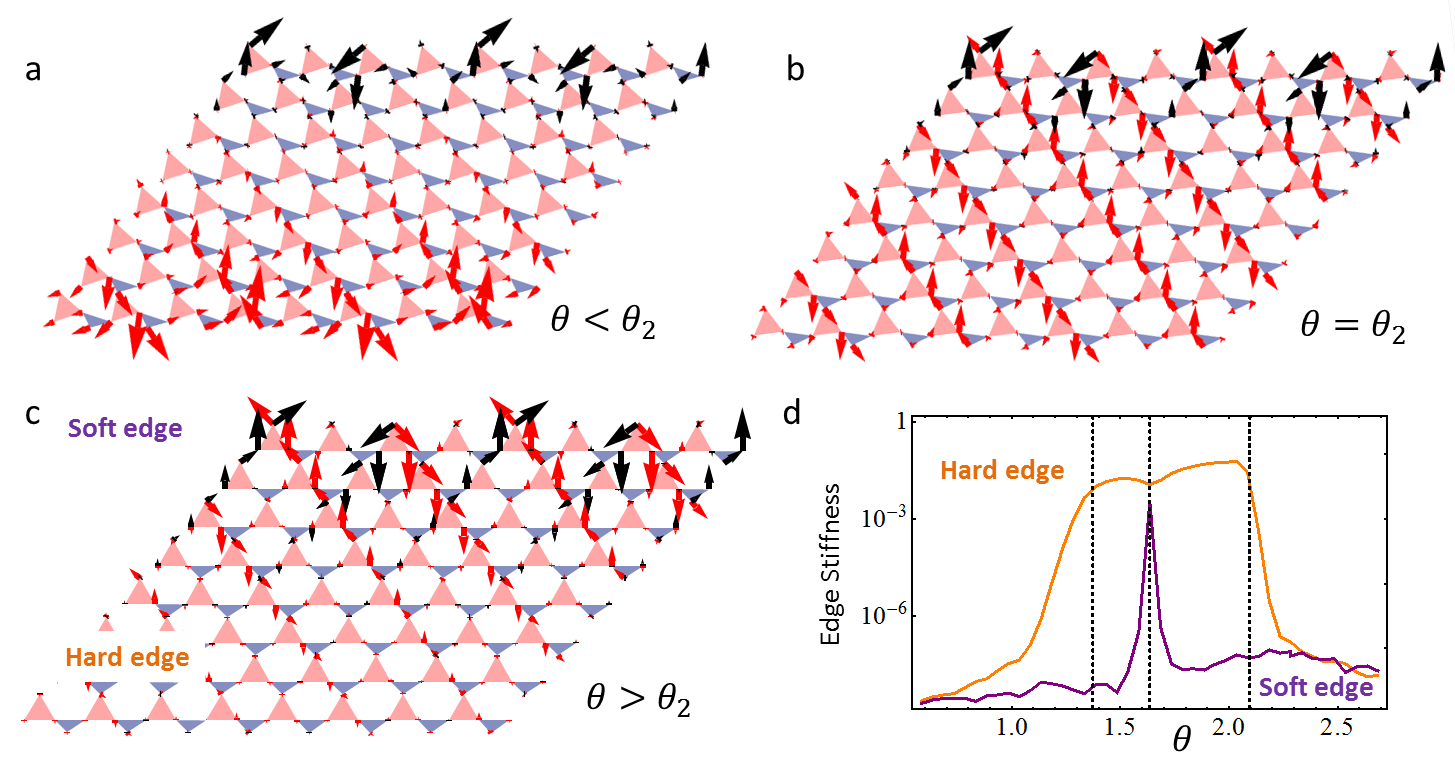}
\caption{(a-c) show the evolution of a pair of floppy modes (red and black arrows) as the example deformed kagome lattice shown in Fig.~\ref{fig:twisting}a traverse its soft twisting coordinate $\theta$ across the critical angle $\theta_2$ where the lattice develops a topological polarization.    Periodic boundary condition is applied to left-right edges and open boundary condition to top-bottom edges. 
(d) Numerical results for the dramatic change of stiffness against local displacements at surfaces as $\theta$ changes, 
 in a 60$\times$60 generic kagome lattice of the structure shown in (a) with free hinges and fixed boundaries except the measurement edge (see Methods).  
}
\label{fig:kagome}
\end{figure*}
\end{center}
\noindent\textbf{Transformations of structures and topological floppy modes}\\
A key result in this work is that the dilation and shear dominant regimes can be realized in the same structure 
and a transition between the two regimes can be achieved by a uniform soft twisting. Here we demonstrate this 
with two examples.  The first example is a deformed kagome lattice~\cite{Kane2014,Grima2005} constructed by connecting rigid triangles 
with free hinges as shown in Fig.~\ref{fig:twisting}a (the term ``deformed'' refers to the fact that this lattice consists of triangles of shapes that differ from those in the regular kagome lattice, and does not mean the lattice is strained).  
%%%%%%%%%%%%%
This structure has one uniform soft twisting, which uniformly rotates  the triangles and varies the angle $\theta$ marked on the figure.
As shown in the \href{http://www-personal.umich.edu/~maox/research/TTMM/TTMM.html}{SI video} , this uniform soft twisting can be readily controlled by a simple expansion of the structure, which transforms the system between dilation dominant and shear dominant regimes, and we further demonstrate dramatic changes in its mechanical properties.

%this structure has one uniform soft twisting which involves rotating uniformly the triangles and thus varying the angle $\theta$ marked on the figure.
As we vary $\theta$, the system goes through five transitions at critical angles $\theta=\theta_1$, $\theta_2$, $\ldots, \theta_5$ respectively (Fig.~\ref{fig:twisting}a).  
For $\theta<\theta_1$ or $\theta>\theta_5$, the system is in the dilation dominant regime $\D>0$ with a rigid bulk and soft edges.
For $\theta_1<\theta<\theta_5$, the system is in the shear dominant regime $\D<0$, where sound modes in the bulk along two directions %$(1, \lambda_{\pm})$ 
show zero velocity.
The edge modes in the shear dominate regime show different characters for different values of $\theta$, according to which the shear dominant regime can be further 
classified into four sub-regimes, separated by critical angles $\theta_2, \theta_3, \theta_4$.  
For $\theta_1<\theta<\theta_2$, floppy modes exist on all edges of the system. 
As $\theta \to \theta_2^{-}$ edge modes on the bottom edge penetrate deeper and deeper into the bulk and eventually become bulk modes (with zero decay rate $\kappa$) at the $\theta = \theta_2$.  Upon further increasing $\theta$, these modes transform into edge modes on the top edge, doubling the number of floppy modes at the top edge. This evolution of floppy modes as $\theta$ increases is illustrated in Fig.~\ref{fig:kagome}a-c.
%As the lattice enters the second sub-regime $\theta_2<\theta<\theta_3$ by increasing $\theta$, the bottom edge loses its floppy edge modes and becomes rigid. These floppy modes are transferred to the top edge, doubling the number of floppy modes at the top edge.  This evolution of floppy modes as $\theta$ increases is illustrated in Fig.~\ref{fig:kagome}a-c.  
%%%%%%%%%%%%%%%%%%%%%%%%%%
%As the system approaches the transition from below $\theta<\theta_2$, edge modes on the bottom edge penetrates deeper and deeper into the bulk and eventually become bulk modes (with zero decay rate $\kappa$) at the transition. Upon further increasing $\theta$, the modes transforms into edge modes on the opposite edge.
%%%%%%%%%%%%%%%%%%%%%%%%%%%%
The transitions at $\theta_3$ and $\theta_4$ are of the same nature, 
where floppy modes shift from certain edges to edges on the opposite side of the system.
These transitions lead to a dramatic change in the edge stiffness.  
We perform conjugate-gradient minimization calculations of the response to a point force on one edge of a lattice with other edges held fixed, and find that the edge stiffness increases by orders of magnitude as floppy modes leave the edge (Fig.~\ref{fig:kagome}d).

The edge properties in these four sub-regimes are dictated by the topological structure of the phonon band, which
is characterized by a vector topological index called ``topological polarization'' ($\mathbf{R}_T$). 
As first discovered in Ref.~\cite{Kane2014}, this topological index points to an edge that gains extra floppy edge modes.
For the deformed kagome lattice discussed above, as $\theta$ crosses the three critical angles  $\theta_2, \theta_3, \theta_4$, the change of $\mathbf{R}_T$ follows $0 \to (\mathbf{a}_2-\mathbf{a}_1) \to \mathbf{a}_2 \to 0$ (where $\mathbf{a}_1$ and $\mathbf{a}_2$ are  the unit vectors of the lattice marked in Fig.~\ref{fig:twisting}a)
%because these four regimes have different $\mathbf{R}_T$
, so the two regimes $\theta_2<\theta<\theta_3$ and $\theta_3<\theta<\theta_4$ have nonzero $\mathbf{R}_T$ (called topologically polarized) and stiff edges in the direction of $-\mathbf{R}_T$.
%The floppy edge modes have to shift from edges to edges across the transition as dictated by the topological index.

The transitions at $\theta_2, \theta_3, \theta_4$ are called topological transitions, because a topological index changes its value across the transitions.
Topological transitions have been well studied for topological states in electronic systems. 
%For topological mechanical systems, 
%although the existence of such a transition is expected, a careful study has been absent to our best knowledge and 
Our design of the \TTMMab\ offers 
a concrete platform to explore these topological transitions in mechanical systems.
%In our system,
%as the system approach the topological transition at $\theta_2$ from below, $\theta<\theta_2$, the edge modes
%along the bottom edge penetrates deeper into the bulk. 
%At $\theta_2$, the penetration depth diverges at $\theta=\theta_2$ 
%(i.e. the decay rate $\kappa\to 0$) indicating that the floppy edge  modes turn into bulk ones. 
At the transition, edges of the triangles form straight lines along certain direction, which is intimately related to the arise of bulk soft modes at the transition. As discussed in Ref~\cite{Sun2012,Kane2014,Lubensky2015}, 
straight lines in the bulk allow states of self stress (possible ways to distribute of internal stress without net forces on any parts) %as well as 
such that floppy bulk  modes %to 
can arise.
% in the system (Fig.~\ref{fig:kagome}a-c).
%If we further increase $\theta$, the bulk zero modes turns into edge modes but along the top edge now for $\theta>\theta_2$. 
%The other two topological transitions $\theta_3$ and $\theta_4$ show the same
%behavior.

As another example, we construct a deformed square lattice using free hinges to connect rigid struts with different lengths (Fig.~\ref{fig:twisting}b).
This structure also has one uniform soft twisting, which changes the angle $\theta$ uniformly. As $\theta$ increases,
the system undergoes one transition from the dilation dominant regime to the shear dominant one.
Agreeing with our elastic theory, the dilation dominant regime shows a rigid bulk and soft edges, while the shear dominant regime has floppy bulk  modes.  
% along two special directions in agreement with prediction from the general elastic theory above, $(1, \lambda_{\pm})$.  
Interestingly, in contrast to the deformed kagome lattice, the deformed square lattice shows no  floppy edge  modes.  Instead it 
has bulk modes with exactly zero energy.  These floppy bulk  modes follow the predicted directions $(1, \lambda_{\pm})$ at small $k$, but deviate at larger $k$ (zero frequency lines in Fig.~\ref{fig:kagome}b are curved).

We emphasize that both of these two examples satisfy the Maxwell lattice condition $\langle z \rangle =2d$ (the deformed kagome lattice is a strut-hinge frame with $z=4$, see Caption of Fig.~\ref{fig:table}), and as a result floppy modes in these structures, either edge modes or bulk modes, are of exactly zero energy, although the general elastic theory discussed above only requires the modes to be soft (i.e., elastic energy scales as $\epsilon^3$ or higher).

%We emphasize that in these two examples, floppy modes have exactly zero elastic energy, although the
%general elastic theory discussed above only requires the modes to be soft 
%(i.e., elastic energy scales as $\epsilon^3$ or higher, instead of $\epsilon^2$). 
%Such exact zero elastic energy is the direct consequence of a counting argument, i.e., the generalized Maxwell's counting rule~\cite{Maxwell1864,Calladine1978} [See Methods for details].

%Finally, very simple design rules for \TTMMab\ stem from our general elastic theory and the Maxwell's counting rule.  
Finally, we provide one design principle, which can be used to generate \TTMMab\  with many different structures.  As shown in Refs.~\cite{Guest2003,Lubensky2015}, 2D structures satisfying $\langle z \rangle =2d$ must have at least one uniform soft twisting (see Methods).  This explains why the deformed kagome and the deformed square lattices we discussed above have uniform soft twistings even with arbitrarily chosen shapes of triangles and strut lengths.  In contrast, the deformed checkerboard lattice (Fig.~\ref{fig:table}) has $z=5>2d$ and the uniform soft twisting disappear when the shape of the parallelograms are changed into arbitrary quadrilaterals.  
Thus by choosing periodic structures with balanced degrees of freedom and constraints ($\langle z \rangle =2d$ in the language of strut-hinge frames) the uniform soft twistings are guaranteed to exist and the structures exhibit zero energy floppy modes.  On the other hand, over-constrained structures with carefully chosen geometry (e.g., the deformed checkerboard lattice) can also exhibit uniform soft twistings but their floppy modes in general are not of zero energy.

A primary challenge in the fabrication of TTMMs is generating sufficiently flexible ``hinges''. The rigidity of rotations at the hinges must be much smaller than the rigidity of deforming the building blocks. When this hinge rigidity is small but finite it determines properties that would vanish with completely flexible hinges, such as the stiffness of the soft edges and sound velocities  of floppy modes.  In this perspective, self-assembly may offer a promising approach.  If tip-to-tip attraction between polygon (colloidal/nano) particles can be realized, an extended periodic 2D lattice may be self-assembled.  Although the tip-to-tip attraction needs to be directional to ensure stability of the open structure, it can be much softer than actually deforming the particles.  Thus binding sites at the tips can serve as flexible hinges for the assembled \TTMMab.

\noindent\textbf{Acknowledgments}\\
We thank Tom C. Lubensky and Vincenzo Vitelli for useful discussions.  DZR thanks NWO and the Delta Institute of Theoretical Physics for supporting his stay at the Institute Lorentz.  This work was supported in part by the ICAM postdoctoral fellowship (DZR) and the National Science Foundation, under grants PHY-1402971 at the University of Michigan (KS).

\textbf{Methods}

\emph{\textbf{Generalized Maxwell's counting rule and uniform soft twistings}}\\
The number of zero modes (modes of deformation which cost no energy) $\numzero$ of a structure is determined by the numbers of degrees of freedom $\numdof$, constraints $\numconst$ and states of self stress (i.e., possible ways to distribute internal stress without net forces on any parts) $\numss$ through the generalized Maxwell's counting rule~\cite{Maxwell1864,Calladine1978}
\begin{align}\label{EQ:Maxwell}
	\numzero = \numdof - \numconst + \numss .
\end{align}
One simple setup to demonstrate this relation is a frame consisting of $N_c$ struts connected at $N$ free hinges (e.g., the structure in Fig.~\ref{fig:twisting}b).  
For a system with spatial dimension $d$, each hinge needs a $d$-component coordinate to describe its location, so it has $d$ degrees of freedom and $\numdof=Nd$.  Each strut fixes the distance between two hinges and thus enforces one constraint.  
It is worthwhile to note that the constraints enforced by struts may not be independent, i.e., some of the struts may be redundant and thus do not introduce 
new constraints. 
As shown in Ref.~\cite{Calladine1978}, each redundant constraint contributes one state of self-stress (i.e., stress may be introduced if the length of the strut change), which is the last term in Eq.~\eqref{EQ:Maxwell}.
The term \emph{isostatic} refers to the special marginal state where $\numzero =d(d+1)/2$ (only trivial zero modes corresponding to rigid translations and rotations of the whole system exist) and $\numss = 0$ where the structure is both stable and stress-free.  
A critical mean coordination number $\langle z \rangle=2d$ for isostaticity~\cite{Thorpe1983,Liu2010,Mao2010} follows from $\numdof = \numconst$, which is a weaker condition of mechanical stability that assumes all struts are independent.  Following the nomenclature of Ref.~\cite{Lubensky2015} we call periodic lattices with $\langle z \rangle=2d$ ``Maxwell lattices''.

%\emph{\textbf{Uniform soft twisting in $z=2d$ lattices}}\\
When the generalized Maxwell's counting rule is applied to \emph{periodic} lattices, as shown in Refs.~\cite{Guest2003,Lubensky2015}, an interesting consequence follows that all lattices with $\langle z \rangle=2d$ (Maxwell lattices) \emph{must} have $d(d-1)/2$ \emph{homogeneous  deformations that are of zero energy}.  For 2D lattices, the case this Article is mainly concerned with, Maxwell lattices have at least one such soft deformation (which we name the uniform soft twisting).  These floppy modes have also been called ``Guest modes''~\cite{Guest2003,Lubensky2015}.

%In order to use this counting rule to examine the existence of uniform soft twisting modes one need to write it in a version based on the repeating unit cells.  This was first done in Ref.~\cite{Guest2003} and more generally discussed in Ref.~\cite{Lubensky2015}.  For our purpose of analyzing uniform soft twisting of periodic structures we can write the number of floppy modes that deform each unit cell in the same way, allowing homogeneous deformations of the whole lattice as
%\begin{align}
%	n_0 = n_{\textrm{d.o.f.}} - n_c + n_{ss} +d^2 ,
%\end{align}
%where the lower case $n$ represent the counting for a unit cell [subscripts defined the same way as in Eq.~\eqref{EQ:Maxwell}], and the extra $d^2$ term represent the extra degrees of freedom from varying the $d$ primitive vectors of the lattice in $d$ dimensions (changing the shape of the unit cell).  Maxwell lattices have $n_c = n_{\textrm{d.o.f.}}$ and thus $n_0 \ge d^2$.  Among these floppy modes $d(d+1)/2$ are trivial rigid translations and rotations of the whole lattice, and thus there are at least $d(d-1)/2$ floppy deformations which are \emph{nontrivial homogeneous elastic deformations}.  For 2D lattices, the case this Article is mainly concerned with, Maxwell lattices have at least one such soft deformation (which we name the uniform soft twisting).  These floppy modes have also been called ``Guest modes''~\cite{Guest2003,Lubensky2015}.

Certain lattices with $\langle z \rangle >2d$, such as the deformed checkerboard lattice in Fig.~\ref{fig:table}, also possess uniform soft twistings, with these necessarily accompanied by states of self stress.

%Even for lattices with $\langle z \rangle >2d$ the existence of states of self stress could also allow such floppy deformations (the deformed checkerboard lattice in Fig.~\ref{fig:table} is an example).
  
%Linearized version of the soft twisting has been named ``Guest modes'' in Ref.~\cite{Lubensky2015}.    
In addition, this type of counting rules and the resulting floppy deformations apply equally to simple frames with struts-hinges and more complicated structures, provided that the degrees of freedom and constraints are countable.  For example, a sub-class of these floppy deformations, the ``rigid-unit-modes'' (RUMs), has been studied in the context of crystals with the structure of periodic corner-touching polyhedra and argued to be responsible for negative thermal expansion in some crystals~\cite{Hammonds1996,Evans1999}, as well as utilized to realize negative Poisson's ratio metamaterials~\cite{Greaves2011,Gatt2015}.  In this Article we discuss more general situations which do not necessarily involve rigid polyhedra.

\emph{\textbf{Numerical calculation of edge stiffness}}\\
Systems of 60 $\times$ 60 unit cells were generated. Three of the four sides were held fixed, while one triangle from the free side was pressed into the structure in the linear regime (qualitatively similar behavior was observed under nonlinear deformations). The Conjugate Gradient method was used to obtain the minimum-energy configuration and the ratio of force to displacement was extracted as the edge stiffness. 
Units were chosen such that the spring constant of the struts and the length of the strut that is horizontal in Fig. 1a were both unity. 

The residual edge stiffness of the soft edge is due to finite size effects as the sides of the lattice are clamped.  
Because the zero modes are exponentially localized to the soft edge, the stiffness of this edge falls exponentially with system size. 
%We conjecture that this stiffness vanish in the infinite volume limit for the ideal system of completely flexible hinges.  
In real systems this soft edge stiffness will be controlled by friction or bending stiffness at the hinges.  
In addition, the sharp rise in the edge stiffness of the soft edge at $\theta_3$ is due to the fine-tuned geometrical effect of the line of struts being pulled taut in the transverse direction.

\appendix

\section{Elastic deformations and the strain tensor}
In order to provide a self-contained discussion, here we first briefly review some basic concepts on elasticity.

In an elastic system, if we focus on macroscopic phenomena at length scales much longer than the scale of the microscopic structure, we can ignore microscopic details and treat the system as a continuous medium. 
In such a picture, each point in the elastic medium can be labeled by its coordinate $\mathbf{r}$  (here we use bold symbols to represent vectors and tensors). Under deformation, the point $\mathbf{r}$ is now displaced to a new location with coordinate $\mathbf{R}$.
Such a deformation is described by a mapping  $\mathbf{r}\rightarrow \mathbf{R}(\mathbf{r})$. In this language, the space that $\mathbf{r}$ lives in is called the \emph{reference space}, i.e., the space before the deformation. 
and the space that $\mathbf{R}$ lives in is dubbed the \emph{target space}, i.e. the space after deformation.

For a slowly varying displacement field, one can keep only the first order derivative $\partial_{i}R_j=\frac{\partial R_j}{\partial r_i}$ (where $i,j$ are Cartesian indices denoting $x,y$ in 2D) in the elastic energy and ignore higher order derivatives.  
%For small deformations, this mapping  $\mathbf{R}(\mathbf{r})$ is dominated by the first (leading) order deriviative , i.e. where the subindicies $i$ and $j$  run over $x$ and $y$ for a 2D elastic mediaum or $x$, $y$ and $z$ in 3D. 
This derivative, $\partial_{i}R_j$, appears to be a rank-2 tensor. However, it is important to realize that the two indices of this matrix live in two different spaces. 
The index $i$ is from $\mathbf{r}$, which lives in the reference space, but the other index $j$ is from $\mathbf{R}$, which lives in the target space.
Symmetry transformations are independent in these two spaces (e.g., a rotation before deformation and the same rotation after deformation result in different strains of the elastic medium).  To express the strain field as a true tensor one can contract either the reference space or the target space indices.  A convenient choice is the \emph{metric tensor}
%To avoid this complication, it is often more conivinient to use a different matrix to describe the elastic deformation, whose two indices live in the same space. Such an objective can be realized utilizing the metric tensor.
\begin{align}
g_{ij}=\partial_{i}R_k \partial_{j}R_k ,
\end{align}
which is a tensor that lives in the reference space ($i,j$ here are both indices in the reference space, and indices in the target space are contracted).  
Here we follow the Einstein summation convention, i.e. the repeated index $k$ is summed over. 
%For the metric tensor $g_{i,}$, both the two indices are from $\mathbf{r}$ and thus they are both indices for the reference space.

It is easy to verify if there is no deformation, $\mathbf{R}(\mathbf{r})=\mathbf{r}$ up to rigid translations and rotations, the metric tensor is the identity matrix.  To describe the strain, the \emph{left Cauchy Green strain tensor} is defined by subtracting the identity matrix from the metric tensor, 
\begin{align}\label{app:eq:epsilong}
\epsilon_{ij}=\frac{1}{2}(g_{ij}-\delta_{ij}),
\end{align}
where $\delta$ represents an identity matrix ($\delta_{ij}=1$ for $i=j$ and $\delta_{ij}=0$ otherwise).

\section{Elastic energy and zero energy deformations}
In this section, we prove that if there exists one uniform deformation that does not cost any elastic energy, the system must also support a series of spatially varying zero-energy deformations.

In general, the energy cost for a elastic deformation, i.e., the elastic energy, is a functional of the strain tensor. To the leading order, the elastic energy is
\begin{align}
E=\int  \mathbf{d r} \; c_{ijkl} \epsilon_{ij}(\mathbf{r}) \epsilon_{kl}(\mathbf{r}) ,
\label{app:eq:elastic_energy}
\end{align}
where $c_{ijkl}$ are elastic constants.  %Effects of higher order terms will be discussed in Sec.~\ref{sec:HOT}.  
We have assumed that the elastic medium has no internal stress.  
This form for elastic energy is a standard description for an elastic medium. For an isotropic medium, these elastic constants reduces into two independent ones, 
bulk and shear moduli. Here, because we are considering a generic system, 
we will maintain this general form and allow the elastic constants to be independent. Same as above, here we adopt the Einstein summation convention,
so all repeated indices are summed over.
The higher order terms, which are not shown in Eq.~\eqref{app:eq:elastic_energy},  contain both higher order terms of the strain tensor as well as spatial derivatives on the strain tensor. Here, we will first ignore these higher order terms and their contributions will be examined in App.~\ref{sec:HOT}.

If an elastic medium has (at least) one uniform deformation, which can be written as a position-independent strain tensor $\tilde{\boldsymbol{\epsilon}}$, that costs no elastic energy, we have
\begin{align}
E=&\int \mathbf{d r} \; c_{ijkl} \tilde{\epsilon}_{ij} \tilde{\epsilon}_{kl} =0 .
\label{app:eq:uniform_flexible_energy}
\end{align}
Because $\tilde{\boldsymbol{\epsilon}}$ is position independent, this indicates
\begin{align}\label{app:eq:zeroE}
	c_{ijkl} \tilde{\epsilon}_{ij} \tilde{\epsilon}_{kl} = 0.
\end{align}
%Consider a flexible system where there exists (at least) one uniform flexible mode, i.e. one uniform deformation that has zero elastic energy.  Same as all other deformations,  this zero energy deformation is also described by a strain tensor, which we will call $\tilde{\epsilon}$. Because this deformation is uniform, $\tilde{\epsilon}$ is position indepenent and therefore, the elastic energy for this deformation is
%\begin{align}
%E=&\int \mathbf{d r} \; c_{ijkl} \tilde{\epsilon}_{ij} \tilde{\epsilon}_{kl}=V c_{ijkl} \tilde{\epsilon}_{ij} \tilde{\epsilon}_{kl} = 0
%\label{app:eq:uniform_flexible_energy}
%\end{align}
%where $V$ is the volumn of the system. Here we used the fact that $c_{ijkl} \tilde{\epsilon}_{ij} \tilde{\epsilon}_{kl}$ is position independent, andthus the integral over the reference space just provides us the volumn of the system $V$.
%Because this is a flexible mode, the elastic energy is zero, which implies that  $c_{i,j;k,l} \tilde{\epsilon}_{i,j} \tilde{\epsilon}_{k,l}=0$

Next, we search for additional spatially varying zero-energy deformations in this system. It is easy to verify that a deformation described by the following strain tensor
\begin{align}
\epsilon_{ij}(\mathbf{r})= \tilde{\epsilon}_{ij}\fr(\mathbf{r}),
\label{app:eq:inhomo_strain}
\end{align}
where $\fr (\mathbf{r})$ is an arbitrary scalar function, has zero elastic energy,
\begin{align}
E=&\int  \mathbf{d r} \; c_{ijkl} \epsilon_{ij}(\mathbf{r}) \epsilon_{kl}(\mathbf{r}) 
= c_{ijkl} \tilde{\epsilon}_{ij} \tilde{\epsilon}_{kl}\int\mathbf{d r} \;  \fr(\mathbf{r})^2
=0,
\label{app:eq:inhomo_flexible_energy}
\end{align}
where we have used the fact that $c_{ijkl} \tilde{\epsilon}_{ij} \tilde{\epsilon}_{kl}=0$ [Eq.~\eqref{app:eq:zeroE}].

\section{Constraints on the function $\fr(\mathbf{r})$ from curvature}
It is important to point out that although the elastic energy [Eq.~\eqref{app:eq:inhomo_flexible_energy}] vanishes for any arbitrary function 
$\fr(\mathbf{r})$, not every function  $\fr(\mathbf{r})$ corresponds to an elastic deformation.
This is because the strain tensor is not an arbitrary rank-2 tensor. 
According to the definition of the strain tensor, in order to ensure that a strain tensor indeed describes a physical deformation, 
there has to exist a deformation $\mathbf{R}(\mathbf{r})$ such that 
\begin{align}
\epsilon_{ij}(\mathbf{r})=\partial_{i}R_k \partial_{j}R_k-\delta_{ij} ,
\end{align}
is satisfied.  
This condition enforces strong strain constraints on the function $\fr(\mathbf{r})$ and in this section we will find the 
necessary and sufficient condition to guarantee a physical zero-energy deformation.

For this purpose, it is more convenient to use the metric tensor instead, which relates to the strain tensor through Eq.~\eqref{app:eq:epsilong}.  
%which is the strain tensor plus an identify metrix, $g_{ij}=\epsilon_{ij}+\delta_{ij}$. 
%Because the metric tenor and the strain tensor 
%contains exactly the same information, 
The question now translates to finding the criterion, under which a metric tensor corresponds to a real physical 
deformation, i.e. to decided whether or not there exists a deformation $\mathbf{R}(\mathbf{r})$ exist such that 
\begin{align}
g_{ij}(\mathbf{r})=\partial_{i}R_k \partial_{j}R_k 
\end{align}
is satisfied.  
The answer to this question has been revealed in the study of differential geometry, where the same question is known as the problem of flat (local) coordinates.
According to {\it Riemann's Theorem}, the necessary and sufficient condition for the existence of such an $\mathbf{R}(\mathbf{r})$ is that the metric tensor 
must have a zero curvature.
The proof of this statement can be found in literature on Riemannian geometry or differential geometry.  
Here, instead of going through the full proof, we provide a physical picture to demonstrate the origin of this zero curvature condition. 
Because both our reference space and the target space (i.e. the material before and after the elastic deformation)
 are defined in a \emph{flat space}, the mapping between these two spaces, $\mathbf{R}(\mathbf{r})$, must not have any nonzero curvature associated with it. 
Therefore, the metric tensor defined from this mapping must have zero curvature~\cite{Frankel2011}.

To determine the curvature for an arbitrary metric tensor $g_{ij}(\mathbf{r})$, 
we first define the Levi-Civita connection, i.e. the Christoffel symbols, using the derivative of $g_{ij}$, 
\begin{align}
\Gamma_{kij}=\frac{1}{2}(\partial_j g_{ki}+\partial_{i} g_{kj}-\partial_{k}g_{ij}).
\end{align}
Then, by taking another derivative to the Levi-Civita connection,  the Ricci curvature tensor is obtained,
\begin{align}\label{app:eq:Ricci}
R_{ijkl}=\partial_{k} \Gamma_{ilj}-\partial_{l}\Gamma_{ikj}+g^{mn}\Gamma_{ikm}\Gamma_{nlj}
-g^{mn}\Gamma_{ilm}\Gamma_{nkj},
\end{align}
where $g^{mn}$ is the matrix inverse of the metric tensor $g_{ij}$.

For a physical deformation in a flat space, the Ricci curvature tensor must vanish, $R_{ijkl}=0$. For the zero energy 
deformations shown in Eq.~\eqref{app:eq:inhomo_strain}, the corresponding metric tenor is
\begin{align}
g_{i j}(\mathbf{r})= \tilde{\epsilon}_{ij} \fr(\mathbf{r})+\delta_{ij}.
\label{app:eq:inhomo_metric}
\end{align}
In 2D, generically, the function $\fr(\mathbf{r})$ depends on both coordinates $x$ and $y$.
However, the zero curvature condition enforces a constraint on $\fr(\mathbf{r})$. Using Eq.~\eqref{app:eq:Ricci}
it is straightforward to verify that 
the curvature vanishes, if and only if $\fr(\mathbf{r})$ takes one of the following two forms
\begin{align}
\fr(\mathbf{r})=f_{+}(x+\lambda_{+} y)
\end{align}
or
\begin{align}
\fr(\mathbf{r})=f_{-}(x+\lambda_{-} y)
\end{align}
Here, $f_+(s)$ and $f_-(s)$ are arbitrary functions of $s$. and $\lambda_+$  and $\lambda_-$ are two constants
that are determined by the strain tensor of the uniform zero-energy deformation
\begin{align}\label{EQ:lambda}
	\lambda_{+}=(\tilde{\epsilon}_{xy}\pm\sqrt{-\det \tilde{\boldsymbol{\epsilon}}})/\tilde{\epsilon}_{xx},
\\
	\lambda_{-}=(\tilde{\epsilon}_{xy}\pm\sqrt{-\det \tilde{\boldsymbol{\epsilon}}})/\tilde{\epsilon}_{xx},
\end{align}
where $\det \tilde{\boldsymbol{\epsilon}}$ is the determinant of $\tilde{\boldsymbol{\epsilon}}$.  It is worth pointing out that this result is independent of the choice of the coordinate.  If the directions of $x,y$ are chosen differently, $\lambda_{\pm}$ will change accordingly, but the two directions given by $x+\lambda_{\pm} y$ are invariant.

We have shown in Eq.~\eqref{app:eq:inhomo_flexible_energy} that 
these deformations cost no elastic energy.
Because the zero curvature condition is the necessary and sufficient condition which guarantees that the strain tensor defined in  Eq.~\eqref{app:eq:inhomo_strain} corresponds to a physical deformation, we conclude that 
the following spatially varying deformations are all zero energy modes of the system
\begin{align}\label{app:eq:strain}
\epsilon_{ij}(\mathbf{r})&=\tilde{\epsilon}_{ij}f_{+}(x+\lambda_{+} y),\nonumber\\
%\label{app:eq:strain_+}
%\end{align}
%or
%\begin{align}
\epsilon_{ij}(\mathbf{r})&=\tilde{\epsilon}_{ij}f_{-}(x+\lambda_{-} y).
%\label{app:eq:strain_-}
\end{align}
Because we can choose arbitrary $f_{+}$ and $f_{-}$, the number of these zero energy deformations is infinite
in the continuous theory. In a real system, with lattice structure and with finite size, the number of zero modes scales with
the linear size of the system $\sim L/a$, where $L$ is the size of the system and $a$ is the lattice constant.  Thus the number of these zero modes is \emph{sub-extensive}.

In summary, we prove here that for a 2D elastic system, as long as there exists one uniform zero-energy mode, 
which is described by a spatially independent strain tensor $\tilde{\boldsymbol{\epsilon}}$, 
there must exist two families of spatially varying zero-energy modes, as shown in Eq.~\eqref{app:eq:strain}.

\section{Higher order terms in the elastic energy}\label{sec:HOT}
In our analysis above, we ignored higher order terms in the elastic energy.  These higher order terms  involve both higher powers in $\boldsymbol{\epsilon}$ and higher order derivatives, such as $\partial \boldsymbol{\epsilon}$.

In the previous section we solved for modes that have zero elastic energy in the leading order theory.  Restoring contributions from higher order terms, the elastic energy of these modes is
\begin{align}
	E=0+O(\boldsymbol{\epsilon}^3)+O(\partial \boldsymbol{\epsilon}\partial \boldsymbol{\epsilon}),
\end{align}
which is small when the strain is small and slowly varying in space.  Thus, strictly speaking, these zero modes should be called \emph{floppy modes} because they are not necessarily exactly zero energy.

In addition, in the Article, we consider frequencies of plane waves (in the bulk or on the surface) that belong to these two families of floppy modes with wave number $k$.  Our theory then predicts that the frequency of these waves are
\begin{align}
	\omega = O(k^2) .
\end{align}
Ordinary plane waves in stable elastic medium have $\omega = c\,k$, where $c$ is the speed of sound.  In contrast, these floppy modes correspond to plane waves with zero speed of sound.

%all invovle extra spaticial derivitives of $\mathbf{R}(\mathbf{r})$,
%under Fou... transform, they correpsonse to higher order in the wavevector $\mathbf{k}$. Therefore, those higher order terms may produce higher order corrections, but those
%corrections will be higher order  $\mathbf{k}$.
%
%For the elastic energy of these deformations, if we ignore the higher order terms, we have proved that the energy is exactly zero. For example, as shown in the Article, if we consider
%a plane-wave (in the bulk or on the surface) zero energy deformation with wavevector $k$, the leading order theory predict zero elastic energy. 
%\begin{align}
%E=0
%\end{align}
%When higher order terms are taken into account, because those term has higher order of $\mathbf{k}$, they can only produce higher order corrections to the elastic energy of order $k^4$.
%\begin{align}
%E=0+O(k^4).
%\end{align}
%If we only focus on long wave length properities, e.g. deformations with long wave length, such a correction will be neglible and invivisble.
%
%For accoustic properties of an elastic media, we usually only focuses on the long wavel length (i.e. small $k$) limit. For example, the definition of the accoustic sound velocity is defined for 
%long wave length sound waves with $k$ near zero. Here, once again, higher order terms can be ignored. And it is well known that only the $O(k^2)$ term in the elastic energy contributes to the sound velicoty.
%For our flexible models, because the elasctic enrgy is zero up to order $O(k^2)$, the sound velocity for these modes must be zero.
This zero sound velocity is a key signature of the systems with floppy uniform deformations that we study here. Regardless of the details of the system, these conclusions hold universally.

In special families of structures with uniform floppy twisting (e.g., Maxwell lattices), these floppy modes may have exactly zero elastic energy, even if higher order terms are taken into account.  This phenomenon is discussed in our Article, where we show that the exact zero elastic energy is protected by Maxwell's counting rule.
%, e.g. isostatic systems where the counting arguments requires the modes to 
%have exact zero energy. 
Nevertheless, it is worthwhile to emphasize that although in the general case (where there is no protection from the counting rule) the elastic energy receives higher order corrections, 
%our conclusion applies to more generic situation, at or away from isostaticity, and in general, the elastic energy will not be eactly zero due to higher 
%order corrections,
the acoustic sound velocity for these modes will always be zero.

\section{Additional information on the \href{http://www-personal.umich.edu/~maox/research/TTMM/TTMM.html}{SI Video}}
\begin{center}
\begin{figure}[h]
\includegraphics[width=.5\textwidth]{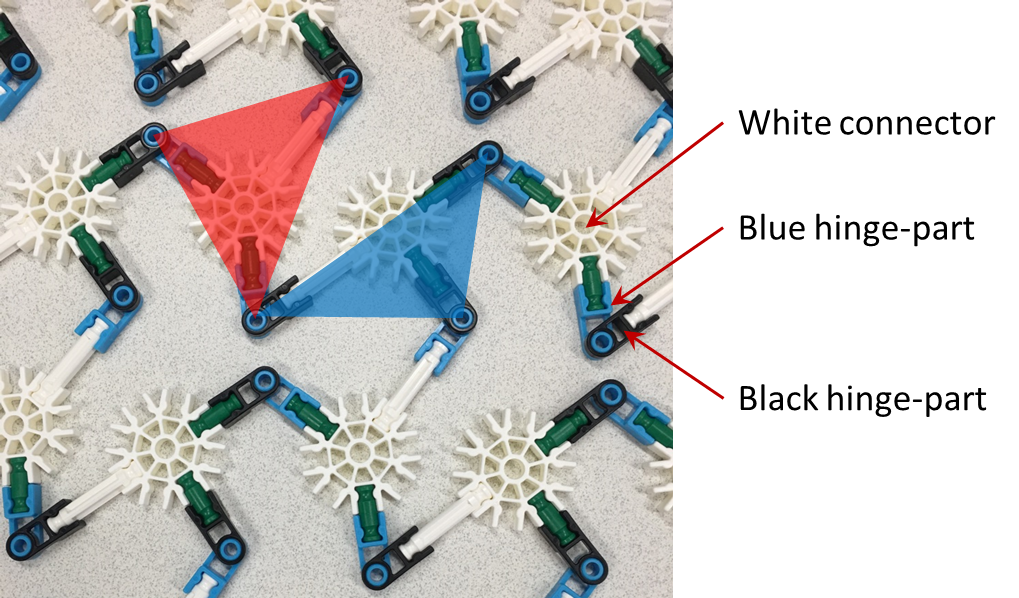}
\caption{
Illustration of the plastic prototype used in the \href{http://www-personal.umich.edu/~maox/research/TTMM/TTMM.html}{SI Video} .
}
\label{fig:knex}
\end{figure}
\end{center}

The prototype is constructed of commercially available plastic "K'NEX" parts. A rigid triangle consists of three rods extending from a central white connector. There are two species of triangles of different shapes (red and blue, as shown in Fig.~\ref{fig:knex}): those ending in blue hinge-parts and those ending in black hinge-parts. Note that although there are no direct connections between two hinge-parts in the same triangle the length between them is fixed by the rods joining them to the central part (which cannot rotate relative to one another) so the triangles are rigid. Each pair of blue hinge-part and black hinge-part form one flexible hinge.  Connected triangles are thus able to rotate freely relative to one another. This is a realization of the \emph{deformed kagome} lattice described in the main text.

The frame consists of four metal rods connected to triangles on the edge of the structure and manipulated by hand. The triangles are free to slide along the lengths of the rods so that the spacing between edge triangles changes even as they remain collinear. The rods are rotated relative to one another, resulting in a uniform soft twisting as described in the main text that alters the lattice structure of the prototype.

%\bibliography{isostaticity3}
\end{document}